\documentclass[twoside,fleqn]{article}
\usepackage{espcrc2}
\usepackage{epsfig}

\title{Flavor Physics and the CKM Matrix:  An Overview}

\author{Adam F.\ Falk \address{Department of Physics and Astronomy, The Johns Hopkins University, 3400 North Charles Street, Baltimore, Maryland 21218 USA}}

\begin{document}
\begin{abstract}
I review the current status of our knowledge of CP violation and flavor physics.  I discuss where one should look for future improvements, and outline the experimental and theoretical priorities of the field.
[Keynote presentation at the Fifth KEK Topical Conference, ``Frontiers in Flavor Physics'', November 20-22, 2001.]
 \end{abstract}

\maketitle

\section{INTRODUCTION}

A decade of studies at LEP, SLC and the Tevatron have taught us a great deal about quarks, leptons, and their interactions.  At the start of the third millenium, we know some important facts with reasonable confidence.  Strong and electroweak physics is described by an $SU(3)\times SU(2)\times U(1)$ Yang-Mills quantum field theory.  The gauge representations of the quarks and leptons are known precisely.  And there are three generations of ``ordinary'' matter, taking the point of view that any neutrino with a mass at the electroweak symmetry breaking scale must be of a different character than the light neutrinos.

The confidence with which we make these statements is not a result of one or two measurements, but rather is the culmination of a long program of precision physics which has tested the nature of electroweak physics richly and with an accuracy at the percent level and below.  In turn, these statements have very specific implications for the structure of flavor changing interactions in the Standard Model.  Such processes depend on a unitary matrix $V_{\rm CKM}$~\cite{Cab}, with the charged current interaction proportional to
\begin{equation}
  V_{\rm CKM}^{ij}\,\overline u_i\gamma^\mu(1-\gamma_5)d_j W_\mu\,.
\end{equation}
If one assumes that only ``ordinary'' matter carries electroweak quantum numbers, then there are three ``up-type'' quarks, $u_i=(u,c,t)$ and three ``down-type'' quarks, $d_i=(d,s,b)$, and $V_{\rm CKM}$ is
\begin{equation}
V_{\rm CKM}=\pmatrix{V_{ud}&V_{us}&V_{ub}\cr V_{cd}&V_{cs}&V_{cb}\cr V_{td}&V_{ts}&V_{tb}}\,.
\end{equation}
Furthermore, tree level neutral current interactions do not change quark flavor.

As a $3\times 3$ unitary matrix, $V_{\rm CKM}$ is fixed by four parameters, one of which is an irreducible complex phase or imaginary part.  The most convenient paramaterization is due to Wolfenstein~\cite{Wolfenstein:1983yz} and is a systematic expansion in the Cabibbo angle,
$\lambda=\sin\theta_C$.  Keeping terms through order $\lambda^3$, $V_{\rm CKM}$ takes the form
$$
\pmatrix{1-\textstyle{1\over2}\lambda^2&\lambda
&A\lambda^3(\rho-{\rm i}\eta)\cr -\lambda&
1-\textstyle{1\over2}\lambda^2&A\lambda^2\cr
A\lambda^3(1-\rho-{\rm 
i}\eta)&-A\lambda^2&1}.
$$
The parameter $\lambda$ is known quite accurately from nuclear $\beta$ decay, $\lambda=0.2196\pm0.0023$~\cite{Groom:in}.  The combination $A\lambda^2=V_{cb}$ is determined from inclusive and exclusive semileptonic $B$ decays.  The single most precise measurement is a recent CLEO analysis~\cite{Cronin-Hennessy:2001fk}, which uses kinematic distributions in inclusive radiative and semileptonic $B$ decays to fix hadronic parameters that are needed for the extraction of $V_{cb}$ from the inclusive semileptonic decay rate.  The result is $V_{cb}=0.0404\pm0.0013$, with uncertainties which are primarily theoretical.  On the other hand, LEP, CLEO and Belle analyses of the exclusive decay $B\to D^*\ell\nu$ yield mutually inconsistent values of $V_{cb}$, so it is not yet possible to compare $V_{cb}$ extracted by the two methods~\cite{vcbexc}.  This is important, because the inclusive anlaysis rests on parton-hadron duality, for which incisive checks are not yet available.  For the moment, it is more conservative to assume a somewhat larger uncertainty than the recent CLEO result, and take $V_{cb}=0.040\pm0.002$.  Further improvement in $V_{cb}$, to percent level precision, is very important.  Lattice computations of the zero recoil form factor for $B\to D^*\ell\nu$ have shown recent progress~\cite{Hashimoto:2001nb}, and future unquenched analyses may be valuable in reducing the uncertainty in $V_{cb}$.

\begin{figure}
\epsfxsize5cm
\epsfbox{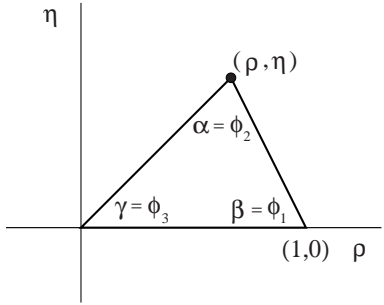}\vskip-0.75cm
\caption{The Unitarity Triangle.}
\label{fig:ut}
\end{figure}

The other two parameters, $\rho$ and $\eta$, are more poorly known.  They may be plotted in the complex plane to give the ``Unitarity Triangle,'' shown in Fig.~\ref{fig:ut}.
The angles $(\phi_1,\phi_2,\phi_3)$ are defined west of the International Date Line, while equivalently $(\beta,\alpha,\gamma)$ are defined to the east  The parameters $\rho$ and $\eta$ enter the nontrivial unitarity relationship between the smallest elements of $V_{\rm CKM}$,
\begin{equation}\label{relation}
  -{V_{td}\over V_{ts}}+{V_{ub}^*\over V_{cb}}=\sin\theta_C\,.
\end{equation}
Verifying this equation would establish the basic structure of $V_{\rm CKM}$. To do so, one must measure independently the magnitude and phase of $V_{td}/V_{ts}$ and $V_{ub}/V_{cb}$.  This is the first important goal of the flavor physics program.

\section{STATUS OF THE CKM MATRIX}

Our current knowledge of the CKM matrix comes from $K$ and, primarily, $B$ physics.  In almost all cases, the experiments are more precise than the theoretical analyses required to extract information on $\rho$ and $\eta$.    Unfortunately, the fact that the dominant errors are theoretical makes it problematic to combine the various measurements in a statistically meaningful way.

\subsection{Information from $K$ physics}

The first experiment to probe the nontrivial $3\times 3$ nature of the CKM matrix was the observation of CP violation in neutral kaon mixing.  The CP violating quantity $\epsilon$ is measured very precisely from the asymmetry between $K_L\to\pi^+\ell^-\bar\nu$ and $K_L\to\pi^-\ell^+\nu$~\cite{Groom:in},
\begin{equation}
  {2{\rm Re}\,\epsilon\over1+|\epsilon|^2}=(3.327\pm0.012)\times 10^{-3}.
\end{equation}
To relate $\epsilon$ to $\rho$ and $\eta$, however, requires one to know the hadronic matrix element
\begin{equation}\label{kbag}
  \langle\overline K{}^0|\,\bar s_L\gamma^\mu d_L\,\bar
  s_L\gamma_\mu d_L\,|K^0\rangle={2\over3}m^2_Kf_K^2 B_K\,.
\end{equation}
Lattice calculations report the bag factor $B_K=0.87\pm0.15$~\cite{Kilcup:1997ye}, but so far only in the quenched approximation.

There is also now conclusive evidence of CP violation in $K_L$ decay.  Combining measurements from KTeV and NA48~\cite{epsp}, we have
\begin{equation}
  {\rm Re}(\epsilon'/\epsilon)= (17.2\pm1.8)\times 10^{-4}.
\end{equation}
This observation is sufficient to rule out ``superweak'' explanations of CP violation.  However, the strong cancelation between contributions to $\epsilon'/\epsilon$ from QCD and electroweak penguins makes the dependence on hadronic physics too sensitive for this measurement to provide useful information on $\rho$ and $\eta$.  Other CP violating kaon physics, such as the analysis of kinematic properties of $K_L\to\pi^+\pi^-e^+e^-$ and $K^+\to\pi^0\mu^+\nu_\mu$, is even harder to interpret theoretically.

\subsection{Information from $B$ physics}

Studies of bottom mesons provide considerably more information about the CKM matrix than do kaons.  The least ambiguous information (but not the most precise) comes from rare semileptonic $B$ decay, a tree level process whose amplitude satisfies
\begin{equation}
  |{\cal A}(b\to u\ell\nu)|^2 \propto |V_{ub}|^2=A^2\lambda^6(\rho^2+\eta^2)\,.
\end{equation}
The magnitude of $|V_{ub}|$ can be extracted from inclusive or exclusive decays; either way, the uncertainties are dominated by theoretical issues. In the case of exclusive transitions such as $B\to\pi\ell\nu$, one needs to compute certain hadronic form factors.  For inclusive decays, rejecting the hundredfold background from $B\to X_c\ell\nu$ requires kinematic cuts which impact severely the model independence of the theoretical analysis.  As a consequence, the best current estimate, $|V_{ub}/V_{cb}|=0.09\pm0.02$~\cite{Ligeti:1999yc}, has a large and poorly understood uncertainty.

The other $B$ physics constraints on the CKM matrix come from loop processes.  Both $B_d$ and $B_s$ mixing are mediated predominantly by fluctuations to intermediate states with $t$ quarks, and so $\Delta M_{B_d}\propto |V_{td}|^2$ and $\Delta M_{B_s}\propto |V_{ts}|^2$.  The experimental world average, $\Delta M_{B_d}=[0.489\pm0.008]\,{\rm ps}^{-1}$~\cite{HameldeMonchenault:2001ae}, is already quite precise, and the limit $\Delta M_{B_s}>14.9\,{\rm ps}^{-1}$~\cite{Groom:in,Convery:2001kg} continues to improve.  To interpret either of these individually requires hadronic physics analogous to the matrix element~(\ref{kbag}); the quenched lattice result is $f_B\sqrt{B_B}\simeq (230\pm 30)\,{\rm MeV}$~\cite{Ryan:2001ej}.  However, most hadronic uncertainties cancel in the ratio,
\begin{equation}
  \Delta M_{B_d}/\Delta M_{B_s}=\xi^{-2} |V_{td}/V_{ts}|^2\,,
\end{equation}
where $SU(3)$ violation is encoded in a parameter $\xi$, for which quenched lattice calculations yield $\xi\simeq1.16\pm0.06$~\cite{Ryan:2001ej}.

Finally, there is the CP violating asymmetry in $B\to\psi K_S$.  To leading order in $\lambda$ in the Wolfenstein parameterization, the mixing amplitude for $B_d\to \overline{B}_d$ carries weak phase $e^{i\phi_1}$, and the decay amplitudes $B_d\to\psi K_S$ and $\overline{B}_d\to\psi K_S$ carry no weak phase.  It then follows that there is no direct CP violation in the decay, and futhermore, that the time dependent CP violating asymmetry is directly proportional to $\sin2\phi_1$,
\begin{equation}
a_{CP}(t)=-\sin2{\phi_1}\,\sin(\Delta M_{B_d}\, t)\,.
\end{equation}
The experimental world average
\begin{equation}
  \sin2{\phi_1}=\sin2\beta=0.79\pm0.10
\end{equation}
is dominated by measurements from Belle and BaBar~\cite{Aubert:2001nu}.  As a result, we know
\begin{equation}
  \arg[V_{td}/V_{ts}]=-26^\circ\pm6^\circ,
\end{equation}
up to a discrete trigonometric ambiguity.  Most important, the quoted uncertainty is statistical and independent of the theory of hadrons.

\begin{figure}
\epsfxsize7.5cm
\epsfbox{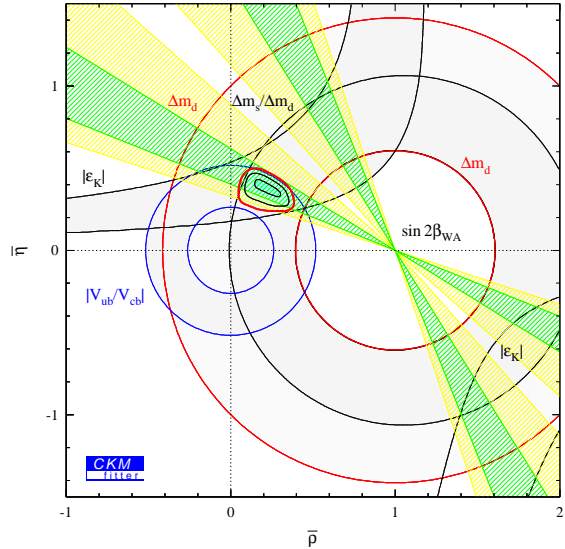}\vskip-1cm
\caption{Status of $V_{\rm CKM}$, from Ref.~\cite{Hocker:2001xe}.}
\label{fig:vckm}
\end{figure}

The current status of our knowledge of $V_{\rm CKM}$ is summarized in Fig.~\ref{fig:vckm}, taken from the CKM Fitter group~\cite{Hocker:2001xe}.
Here the constraints are plotted for $\bar\eta=\eta(1-\lambda^2/2)$ and  $\bar\rho=\rho(1-\lambda^2/2)$, which incorporates certain higher order corrections in $\lambda$.  Note that the agreement of the measurements, while not yet particularly precise, is already quite nontrivial.  While there are not yet any direct constraints on $\phi_3=\gamma$, the global fit seems to prefer $\phi_3<90^\circ$.  It is also interesting to observe that the limit on $\Delta M_{B_s}/\Delta M_{B_d}$ is quite restrictive, and much larger values of $\Delta M_{B_s}$ would be difficult to accomodate simultaneously with the other data.

\section{IMPENDING IMPROVEMENTS}

At present, the analysis of CP violation in $B\to\psi K_S$ provides the only statistically dominated two sided constraint on the Unitarity Triangle.  It will not be possible to exploit the precision of this measurement, however, until the other constraints have been improved to the point that they also come with meaningful uncertainties at the ten percent level or below.  To test the relation~(\ref{relation}), we would like to have this quality of knowledge of the magnitudes and phases of $V_{ub}/V_{cb}$ and $V_{td}/V_{ts}$.  The argument of $V_{td}/V_{ts}$ is now known; in this section I discuss future improvements in the other three quantities.

\subsection{Improvement in $|V_{ub}/V_{cb}|$}

In the next few years, we should see improvements in the extraction of $|V_{ub}|$ from both inclusive and exclusive semileptonic $B$ decays.  In the exclusive case, the fundamental limitation is our knowledge of the hadronic matrix elements $\langle\pi|\bar u_L\gamma^\mu b_L|B\rangle$ and $\langle\rho|\bar u_L\gamma^\mu b_L|B\rangle$.  Soon unquenched lattice calculations will become the standard, at which point we can expect the lattice to provide results for which all uncertainties are under control.  It is fortuitous that the computation of $B\to\pi$ is best performed near zero recoil, where the experiments are also the most sensitive.  (Euclidean space computations of $B\to\rho$ may always remain problematic because of the finite $\rho$ width.)  What remains unclear is how accurate the lattice predictions ultimately will be, not to mention when they will actually come~\cite{El-Khadra:2001rv}.

The situation is more complicated, and perhaps more interesting, for the inclusive case.  Up to the prefactor $|V_{ub}|^2$ itself, one can use Heavy Quark Effective Theory techniques to compute the fully inclusive rate for $B\to X_u\ell\nu$ to approximately ten percent, with the primary uncertainty arising from the value of the $b$ quark mass~\cite{ligetireview}.  Unfortunately, the fully inclusive rate is not something which can be measured experimentally, because of the background from semileptonic decays to charm.  This background can be suppressed only by imposing strict kinematic cuts.  Existing analyses have used either the charged lepton energy,
\begin{equation}
E_\ell >(m_B^2-m_D^2)/2m_B\,,
\end{equation}
or, after reconstructing the energy and momentum of the missing neutrino, the invariant mass of the hadronic final state,
\begin{equation}
m(X_u)<m_D\,.
\end{equation}
Because of uncertainties in the neutrino reconstruction, the actual invariant mass cut is typically somewhat stronger.

The problem is that both of these cuts include events in the ``light cone region'', in which the $u$ quark is near its maximum energy.  In this regime, there are large effects from Sudakov logarithms of the form $\alpha_s^{n}\log^{2n}(1-y)$ as $y\to1$, as well as sensitivity at leading order to the $b$ quark light cone distribution function $f(k^+)$~\cite{NFJMW}.  The Sudakov logarithms lead to large radiative corrections which can, with enough work, be resummed~\cite{Akhoury:1995fp}.  But $f(k^+)$, known as the ``shape function,'' introduces an irreducible model dependence into the extraction of $|V_{ub}|$.  Although the shape function is universal, and can be measured in a process such as $B\to X_s\gamma$ and then applied to $B\to X_u\ell\nu$~\cite{Neubert:1993um}, what has actually been done in practice is either to model the shape function or to fit some ansatz for it simultaneously with extracting $|V_{ub}|$.  In either approach, the theoretical uncertainties are difficulty to quantify meaningfully~\cite{Abreu:2000mx}.

An alternative approach recently has been proposed, which should largely eliminate these problems~\cite{Bauer:2001rc}.  One can cut on the dilepton invariant mass,
\begin{equation}
q^2=(p_e+p_\nu)^2>(m_B-m_D)^2\,,
\end{equation}
a restriction which eliminates the light cone region as well.  The disadvantage is that only $15-20\%$ of the rate is included with a pure $q^2$ cut, raising questions about the applicability of parton-hadron duality.  The best analysis will combine a $q^2$ cut with a cut on $m(X_u)$, in such a way as to include a larger fraction of the total rate but with a substantially reduced sensitivity to shape function and Sudakov effects.

Taking the improvements in exclusive and inclusive analyses together, I would expect that in the next few years we will know $|V_{ub}/V_{cb}|$ to approximately ten percent, with a meaningful uncertainty.

\subsection{Improvement in $|V_{td}/V_{ts}|$}

The uncertainty in the extraction of $|V_{td}|$ from $\Delta M_{B_d}$ is dominated by our imperfect knowledge of the hadronic quantity $f_B\sqrt{B_B}$.  While future unquenched lattice calculations will improve on the current accuracy of about $15\%$, it will be more effective to extract $|V_{td}/V_{ts}|$ directly from the ratio $\Delta M_{B_d}/\Delta M_{B_s}$.  At Tevatron Run II, the present lower limit on $\Delta M_{B_s}$ should become an actual measurement, if its value is anywhere near that predicted by the Standard Model.  CDF is likely to determine $\Delta M_{B_s}$ with better than one percent accuracy.

Then it will up to theory to try to match the quality of experiment.  Although the dominant hadronic uncertainties cancel in the ratio, the lattice community still will need to provide a reliable unquenched calculation of
\begin{equation}
  \xi=({f_{B_s}\sqrt{B_{B_s}})/(f_{B_d}\sqrt{B_{B_s}}}).
\end{equation}
The quenched result is quoted as $\xi=1.16\pm0.06$.  However, some of the first unquenched calculations may be seeing large effects associated with soft pion loops~\cite{Yamada:2001xp}.  These chiral logarithms give a curvature which must be taken into account in the extrapolation $m_q\to0$~\cite{Sharpe:1995qp}.  If present, such a curvature would have the effect of raising $\xi$.  Until this situation is clarified, we will follow Ref.~\cite{Ryan:2001ej} in assigning an additional provisional one sided uncertainty, and quote
\begin{equation}
\xi=1.16^{+0.13}_{-0.06}\,.
\end{equation}
As the unquenched calculations mature and the chiral extrapolation is better understood, the error should be reduced again.

With the experiments already taking data and the anticipated theoretical improvements, I would expect that in the next few years we will know $|V_{td}/V_{ts}|$ to approximately five percent.

\subsection{Improvement in $\arg[V_{ub}/V_{cb}]$}

The most difficult parameter of the Unitarity Triangle to determine accurately is $\phi_3=-\arg[V_{ub}/V_{cb}]$. Consequently, there are many proposals for doing so, lying roughly on a spectrum with {\it theoretically controlled\/} on one end and {\it experimentally feasible\/} on the other.  No single best method has yet been identified, and here I will discuss three representative and popular ones which are available in the $B$ Factory era.

\subsubsection{The clean method}

The theoretically cleanest method for extracting $\phi_3$ is to use CP violating rate asymmetries in decays mediated by the quark transitions $b\to u\bar cs$ and $b\to u\bar cd$.  For example, one can study the interference between the decay chains $B^-\to K^-D^0\to K^-f_i$ and $B^-\to K^-{\overline D}{}^0\to K^-f_i$, by comparing to the CP conjugate $B^+\to K^+\overline{f}_i$ decays~\cite{Atwood:1996ci}.  Since $B^-\to K^-D^0$ is mediated by $b\to c\bar us$, its weak phase is real in the Wolfenstein parameterization, while $B^-\to K^-\overline{D}{}^0$ is mediated by $b\to u\bar cs$ and has an amplitude proportional to $e^{-i\phi_3}$.  Hence the interference is sensitive to $\sin\phi_3$, as well as to strong phase differences which must be extracted simultaneously.  One needs measurements with at least two different final states $f_i=(K^+\pi^-,K\pi\pi,\ldots)$ to eliminate the sensitivity to hadronic physics.

Despite its theoretical attractiveness, there are considerable difficulties in using this method to measure $\sin\phi_3$.  The combined branching fractions are at the level of $10^{-7}$ or smaller, and one needs accurate measurements in many channels.  The method also depends on unknown strong phase differences, which must not be negligible for there to be sensitivity to $\sin\phi_3$.  There are other clean methods which exploit the same quark level transitions, but each  poses practical challenges of its own, such as measuring small CP asymmetries of order $\lambda^2$, or performing geometrical constructions with ``squashed'' amplitude triangles.

\subsubsection{The classic method}

The original proposal to extract $\phi_3$ at the $B$ Factories was to study CP violation in the quark transition $b\to u\bar ud$, by analyzing the time-dependent asymmetry in $B\to\pi^+\pi^-$.  Since $\pi^+\pi^-$ is a CP eigenstate, this asymmetry depends on CP violation in the interference between $B_d$ mixing and decay.  In the Wolfenstein parameterization, the mixing phase is $e^{2i\phi_1}$, so the comparison of $B^0(t)\to\pi^+\pi^-$ with $\overline{B}{}^0(t)\to\pi^+\pi^-$ is sensitive to the combination $\sin(2\gamma^*+2\phi_1)$, where $e^{i\gamma^*}$ is the phase of the decay amplitude.

The difficulty with this method arises in relating $\gamma^*$ to angles in the Unitarity Triangle.  Because of the $\bar uu$ pair in the final state, there are contributions from both penguin and tree level processes, which carry different weak phases.  The amplitude takes the form
\begin{equation}
  {\cal A}(\overline{B}{}^0\to\pi^+\pi^-)=Te^{i\phi_3}+Pe^{-i\phi_1}=Ae^{i\gamma^*},
\end{equation}
where the relation of $\gamma^*$ to $\phi_1$ and $\phi_3$ depends on the ratio $P/T$ of hadronic matrix elements.  In the absence of ``penguin pollution,'' in which case $P=0$, we have $\gamma^*=\phi_3$ and the analysis is sensitive directly to the CKM angle $\alpha=\phi_2=\pi-\phi_1-\phi_3$.  However, there is no reason to neglect penguin diagrams; most experimental and theoretical estimates yield values in the range $|P/T|\sim 0.1-0.5$.

To make further progress, one must have some reliable information about the magnitude and phase of $P/T$.  One possibility is to measure or bound the penguin contribution~\cite{Gronau:1990ka}.  For $B\to\pi\pi$, the magnitude can be estimated by using flavor $SU(3)$ symmetry and the rates for penguin dominated $B\to K\pi$ and $B\to KK$ processes, but the phase is not accessible.  Alternatively, one can use the flavor tagged rates for $B\to\pi^0\pi^0$ to perform an ``isospin analysis,'' which avoids potentially large $SU(3)$ violating corrections.  However, the very rare four photon final state is probably too hard to identify at the $B$ Factories, although an upper bound on the $B\to\pi^0\pi^0$ rate could be useful for constraining $|P/T|$.  An interesting elaboration on the isospin analysis is to study the more complicated $\pi^+\pi^-\pi^0$ final state, resonating through the various $\rho\pi$ channels.  Here the most difficult experimental issues are substracting nonresonant backgrounds and obtaining sufficient statistics to perform a full Dalitz plot analysis.  

Finally, some theorists propose that one can compute $P/T$ using short distance QCD techniques in the limit $m_b\to\infty$.  There are advocates for two approaches, known as ``QCD factorization''~\cite{Beneke:2001ev} and ``perturbative QCD"~\cite{Li:2001vm}.  Starting from quite different assumptions, both approaches give approximately $|P/T|=0.3$.  It is not yet clear whether this agreement is more than accidental.  In any case, the methods require further development before they will be mature enough to be useful for reliable predictions.  In particular, it remains a crucial outstanding problem to organize and understand the full suite of corrections beyond leading order in $1/m_b$.

\subsubsection{The phenomenological method}

Finally, there is by now an extensive literature of proposals to extract information on $\phi_3$ by combining rates and asymmetries in $B\to\pi\pi,K\pi,KK$ transitions~\cite{Fleischer:2001zn}.  These analyses make a virtue of penguin pollution by exploiting the varieties of interference which can be observed in different modes.  Two classic examples are the Fleischer-Mannel bound~\cite{Fleischer:1997um},
\begin{equation}\label{fm}
  \sin^2{\phi_3}\le R\big(1+2{\epsilon}\sqrt{1-R}\big),
\end{equation}
where
\begin{equation}
  R = \frac{{\cal B}(B^0 \to \pi^- K^+) + {\cal B}(\overline{B}{}^0 \to \pi^+ K^-)}{
  {\cal B}(B^+ \to \pi^+ K^0) + {\cal B}(B^- \to \pi^- \overline{K}{}^0)}\,,
\end{equation}
and the Neubert-Rosner bound~\cite{Neubert:1998jq},
\begin{eqnarray}\label{nr}
  R_*^{-1}\le\left(1+{\bar\epsilon_{3/2}}|{ \delta_{\rm EW}}-\cos{\phi_3}|\right)^2
  \hphantom{mm}   \nonumber\\
  \mbox{}+{\bar\epsilon_{3/2}}\left({\bar\epsilon_{3/2}}+2|{\epsilon_a}|\right)\sin^2{\phi_3}\,,
\end{eqnarray}
where
\begin{equation}
  R_* = \frac{{\cal B}(B^+ \to \pi^+ K^0) + {\cal B}(B^- \to \pi^- \overline{K}{}^0)}{{\cal B}(B^+ \to \pi^0    
  K^+) + {\cal B}(B^- \to \pi^0 K^-)}\,.
\end{equation}
There are two difficulties which typically plague such methods.  First, the bounds depend on additional hadronic parameters, such as $\epsilon$, $\epsilon_a$, $\bar\epsilon_{3/2}$ and $\delta_{\rm EW}$ above, whose determination requires theoretical inputs such as flavor $SU(3)$ or nonperturbative dynamics~\cite{Falk:1998wc}.  This intrusion of hadronic physics usually introduces an unwelcome and difficult to quantify model dependence into the extraction of $\phi_3$, although clever analyses do their best to minimize this. Second, the methods often only give information if some observable is found to lie in a particular range; for example, the bounds (\ref{fm}) and (\ref{nr}) require $R,R_*<1$.

To summarize, there are many ideas for improving our knowledge of $\arg[V_{ub}/V_{cb}]$, but none that could be described as a sure bet.  In contrast to $|V_{ub}/V_{cb}|$ and $|V_{td}/V_{ts}|$, it is unclear how precisely we will determine $\phi_3$ in the next few years.

\section{FLAVOR PHYSICS BY  2005}

By the middle of this decade, the first phase of the $B$ Factory era will be mature.  We can expect BaBar and Belle each to have collected as much as $500\,{\rm fb}^{-1}$ of data, and for Tevatron Run II to be well into high luminosity running.  The primary goal of flavor physics in this era is to verify the gross consistency of the CKM picture.  Since the $(\rho,\eta)$ plane is a two dimensional parameter space, the incisiveness of the test will depend on the precision of the third and fourth best measurements.  Most likely, these will be the magnitude and phase of $V_{ub}/V_{cb}$, and I believe we can optimistically hope for a probe of $V_{\rm CKM}$ at the level of five to ten percent.  As exciting and important as the recent measurements of $\sin2\phi_1$ have been, further improvement in this quantity cannot by itself teach us more about the adequacy of the CKM matrix for describing flavor physics.

The next phase of experimental flavor physics depends on whether or not the results of the first phase are consistent with the Standard Model.

The optimistic scenario is that the global set of measurements cannot be explained by the standard CKM picture.  For example, if we were to find $\Delta M_{B_s}=30\,{\rm ps}^{-1}$ and $\phi_3=(110\pm20)^\circ$, the Standard Model interpretation would fail unambiguously.  Then our attention would turn to the crucial question of how this failure should be interpreted.  One possibility would be that there are nonstandard contributions to $b$ quark charged current interactions, such as a right handed $W_R^\mu\,\bar u_R\gamma_\mu b_R$ coupling.  Such models are quite constrained, but not entirely ruled out~\cite{Lewandowski:2000iv}.

However, the most plausible interpretation would be that there are new contributions to one or more processes which arise only at the loop level in the Standard Model.  For example, $B_d$ and $B_s$ mixing come from loops with virtual $t$ and $W$; in supersymmetric extensions, such mixing can be induced by virtual $\tilde t$ and $\widetilde w$ as well.  Unfortunately, it is hard to make a clear prediction for the size of the supersymmetric contribution to $B$ mixing.  In the Standard Model, the mixing amplitude has the structure
\begin{equation}
  \Delta M_{B_q}\sim\, |V_{tb}V_{tq}|^2\cdot{(\Lambda_{\rm QCD}^3/ M_W^2)}\,,
\end{equation}
where the first factor encodes flavor symmetry breaking and the second encodes the suppression from the $B$ meson fluctuating to a virtual state at the electroweak scale.  The same structure applies to supersymmetric contributions.  If $m_{\tilde t}, m_{\widetilde w}<200\,{\rm GeV}$, then the mass suppression will be similar to that in the Standard Model.  But almost nothing is really known about the nature of flavor symmetry breaking in supersymmetric models.  A significant suppression of flavor violation in the first two generations is necessary for consistency with data on neutral $K$ mixing, but the source of this suppression and its implications for the third generation remain unclear.  Without some understanding of the flavor problem, it is impossible to say whether observable contributions to $b$ physics are a generic feature of supersymmetry.

In any case, if the measurements are inconsistent with the Standard Model then there must be new flavor interactions of some kind at or near the electroweak scale, and they will have been revealed first in $B$ physics!  We will learn later what the new physics is, at the Large Hadron Collider and/or at a future linear electron collider.  There are many possibilities beyond supersymmetry, including technicolor, extra generations, left-right models, GUT relic such as $Z'$ bosons, or even extra spatial dimensions or low energy string theory.  It will be an exciting time.

There is also the pessimistic scenario, in which the global set of measurements can be accomodated by the CKM picture, at the ten percent level or so.   Of course, this could happen by accident even in the presence of large new contributions to $B$ processes, especially given the discrete ambiguities in the extraction of the angles.  If the data appear consistent with the Standard Model, it will be very important to resolve these ambiguities.  For this purpose, methods which are  each problematic in isolation will be valuable taken together, because of their sensitivity to different combinations of $\phi_1$ and $\phi_3$.  For example, one can extract $\sin\phi_3$ by combining rates and asymmetries for $B\to\pi\pi,K\pi,KK$, $\sin(2\phi_1+\phi_3)$ from asymmetries in $B\to D\pi$, and $\sin(2\phi_1+2\phi_3)$ from time dependent analyses of $B\to\pi\pi$ and $B\to\rho\pi$.  Even crude measurements will help resolve ambiguities if one precise measurement is also in hand.

Nevertheless, if no inconsistency with the Standard Model is found then the most natural interpretation will be that the CKM matrix has been determined precisely.  This would be a tremendous accomplishment!  But what is the next step?

\section{FLAVOR PHYSICS BEYOND 2005}

In rough terms, the period after 2005 will constitute a second phase of the precision flavor physics program, with priorities determined in part by what the first phase has revealed.  Nonetheless, we already can discern certain key priorities.  It will be important both to push to higher precision and sensitivity in processes accessible to the $B$ Factories, and to develop a robust program of $B_s$ physics.  There will also be a crucial role for experiments in the $K$ and $D$ systems.  I will describe some of the highlights below.

\subsection{Precision and ``redundancy'' in $V_{\rm CKM}$}

Especially if no sign of new physics has appeared, it will be important to increase the precision of the test of the Standard Model.  This precision will depend, most likely, on improvements in $|V_{ub}|$ and $\phi_3$.  Pushing our knowledge of $|V_{ub}|$ to the five percent level will require high statistics and careful, subtle analyses.  The program will benefit particularly to the extent that one is able to loosen the kinematic cuts used to reject $b\to c\ell\nu$.  Absent a new theoretical idea, this will involve hard work and a detailed understanding of the BaBar and Belle detectors.

A clean and precise measurement of $\phi_3$ probably must wait until it is possible to study the time-dependent asymmetry in $B_s\to D_sK$~\cite{Aleksan:1991nh}.  The amplitude for $\overline{B}_s\to D_s^+K^-$ carries the weak phase $e^{i\phi_3}$ in the Wolfenstein parameterization, while $B_s\to D_s^+K^-$ has none.  These transitions receive no penguin contributions, so if there is no CP violation in $B_s$ mxing then the asymmetry is unambiguously sensitive to $\sin\phi_3$.  The accuracy which can be attained will depend on the values of $\Delta M_{B_s}$ and various $B_s$ and $D_s$ branching ratios.  The measurement is most likely to be performed at a hadronic $B$ experiment such as LHCb or BTeV.

It will also be important to perform ``redundant'' measurements of CKM parameters.  Because the CKM picture is so constrained, the amplitudes of many different quark level transitions are correlated.  In the presence of new physics these correlations are broken, and looking for such effects is an important tool.  For example,  one can extract $\phi_3$ from $B_s\to D_sK$, where the decay is mediated by $b\to u\bar cs$, or from $B\to\rho\pi$, which is mediated by $b\to u\bar ud$.  In the Standard Model both amplitudes are proportional to $A\lambda^3 e^{i\phi_3}$, but since new physics could affect either one or both transitions, the comparison between $\phi_3$ measured in the two is important.  Similarly, one should check null predictions of the Standard Model, such as the tiny CP violating asymmetries in $B_s(t)\to\psi\phi,\psi\eta$.  An observation larger than a few percent would be a clear sign of new contributions to $B_s$ mixing.

\subsection{Rare $B$ decays}

For the most part, new physics at the electroweak scale will only contribute significantly to a $B$ decay if it competes with a Standard Model loop process.  Therefore neutral current decays, which do not arise at tree level, have a particular role to play in constraining new physics scenarios.  Because of hadronic uncertainties, one must look for large deviations from standard predictions to search for new effects.  Key progress will come not from more precision in processes which are understood, but from studying ever rarer decays and looking for large effects in subtle properties.

The radiative decay $B\to X_s\gamma$ is in reasonably good agreement with the Standard Model, with both experiment and theory giving a branching fraction of approximately $3\times10^{-4}$~\cite{Chen:2001fj}.  There is no room for a big surprise in this process, nor in the exclusive channel $B\to K^*\gamma$.  Recently, attention has turned to the even rarer mode $B\to X_s\ell^+\ell^-$, for which the inclusive branching fraction is approximately $7\times 10^{-6}$.  What is interesting about this decay is that the Standard Model predicts a particular chiral structure for the quark transition, $\bar s_L\gamma^\mu b_L\cdot\bar\ell\gamma_\mu\ell$.  This structure reveals itself, for example, in the forward-backward lepton asymmetry in $B\to K^*\ell^+\ell^-$~\cite{Burdman:1998mk}.  The upper bound on the branching fraction in this channel is $2.5\times10^{-6}$~\cite{nashlp}, so there is little room for an enhancement in the rate from new physics, but it would still be possible to probe a supersymmetric contribution to an operator of different chirality such as $\bar s_R\gamma^\mu b_R\cdot\bar\ell\gamma_\mu\ell$.  It is also intriguing to note that Belle has  reported a $4\sigma$ signal for $B\to K\ell^+\ell^-$~\cite{bellekll}.

\subsection{Rare $K$ decays}

The very rare process $K\to\pi\nu\bar\nu$ is an excellent mode in which to probe the CKM matrix and search for new flavor physics affecting the first two generations.  In the Standard Model, the transition $s\to d\nu\bar\nu$ is mediated primarily by a $tW$ intermediate state, and therefore is sensitive to the combination $V_{td}V^*_{ts}$.  One can eliminate hadronic uncertainties by normalizing the amplitude to the semileptonic decay,
\begin{equation}
  {{\cal A}({ K\to\pi\nu\bar\nu})\over{\cal A}( K\to \pi\ell\bar\nu)
  }\propto F({\rho},{\eta}; m_q,M_W)\,,
\end{equation}
where the right hand side depends only on particle masses and CKM parameters.

The process can be probed in either the charged or neutral channel, which yield different constraints on the Unitarity Triangle.  The charged decay $K^+\to\pi^+\nu\bar\nu$ has a Standard Model branching ratio of approximately $10^{-10}$. The E787 Collaboration has seen two events and reports~\cite{Adler:2001xv}
\begin{equation}
  {\cal B}(K^+\to\pi^+\nu\bar\nu)=\big[1.57^{+1.75}_{-0.82}\big]\times 10^{-10}\,,
\end{equation}
consistent with expectation but with errors which are still quite large.  The relation between the rate for $K^+\to\pi^+\nu\bar\nu$ and a constraint on $\rho$ and $\eta$ is complicated by the fact that the $cW$ intermediate state contributes as well, so
\begin{equation}
  {\cal B}(K^+\to\pi^+\nu\bar\nu)\sim|{ V_{td}V_{ts}^*}+f(m_c^2){ V_{cd}V_{cs}^*}|^2\,.
\end{equation}
The dominant theoretical uncertainty, which is below the ten percent level, is from the value of $m_c$~\cite{Buchalla:1998ba}.  The CKM Collaboration hopes to measure the rate to $20\%$ at Fermilab by 2010, leading to a $10\%$ constraint on the Unitarity Triangle.

The neutral channel $K_L\to\pi^0\nu\bar\nu$ is rarer by yet another order of magnitude.  It is also harder to observe, due to the absence of any charged particle in the initial or final state.  On the other hand, it is theoretically cleaner in that the transition is almost purely CP violating and therefore receives a negligible contribution from charm.  If both the $K^+$ and $K_L$ decays could be measured at the ten percent level, this would provide a powerful independent determination of the Unitarity Triangle, with no reference to $B$ physics.

\subsection{$D$ meson mixing} 

The mixing of neutral $D$ mesons is very different from the $K$ and $B$ systems, because it is mediated by intermediate down-type quarks.  Since there is no large flavor violation from a virtual $t$ quark, $D$ mixing is suppressed by $SU(3)$ symmetry and is quite small in the Standard Model.  Furthermore, CKM suppressions imply that the process is dominated by $d$ and $s$ intermediate states, so $D$ mixing involves long distance physics in an essential way.  Therefore it is quite difficult to compute in the Standard Model.  The near absence of $b$ virtual states also implies that $D$ mixing is almost exactly CP conserving.

The current experimental situation is somewhat murky.  Four experiments are sensitive to a quantity $y_{CP}$, which is equal to $y=\Delta\Gamma/2\Gamma$ if $D$ mixing conserves CP~\cite{Bergmann:2000id}.  The  results are~\cite{Ligeti:1999yc}
\begin{equation}
  {y_{CP}}=\cases{\hfill  3.4\pm  1.6\%\ &({\rm FOCUS})\cr  -1.1\pm2.9\%\ &({\rm CLEO})\cr
  -1.1\pm2.8\%\ &({\rm BaBar})\cr  -0.5\pm1.2\%\ &({\rm Belle})\cr}
\end{equation}
Although the FOCUS result is intriguing, taken together these results show no sign of nonzero $y$.
For $x=\Delta M_D/\Gamma$, there is not yet even a hint of a signal.  The current experimental upper limit is at the level of a few percent, although the precise bound depends on certain theoretical assumptions~\cite{Godang:1999yd}.

Given that large values of $x$ and $y$ are already ruled out, the question arises whether the existing bounds are plausibly saturated by the Standard Model expectation.  This may well be the case.  Although short distance calculations typically predict $x$ and $y$ order $10^{-3}$ or below~\cite{Georgi:1992as}, such estimates may not be realistic.  A recent analysis of $SU(3)$ violation in the phase space of the intermediate states yields $y\sim10^{-2}$, with $x$ not much smaller~\cite{Falk:2001hx}.  These predictions cannot be relied on in detail, but  they are sufficient to argue that no future measurement of $x$ and $y$ consistent with the existing bounds could give unambiguous evidence of new physics.  The lesson is that  our attention must turn to the search for CP violation in $D$ mixing, the  absence of which in the Standard Model is independent of  hadronic uncertainties.

\section{TOOLS FOR B PHYSICS}

Although the $K$ and $D$ systems have a role to play, it is the rich phenomenology of $B$ mesons that will dominate experimental flavor physics over the next decade.  Fortunately, we have entered an era in which wonderful tools for this purpose have become available, with the prospect of a second generation of experiments to follow.

Of course, the asymmetric $B$ Factories are the stars of the moment.  Belle and BaBar, taking data respectively at KEK-B and PEP-II, can each integrate more than $0.25\,{\rm fb}^{-1}$ in a single day.  By comparison, the entire CLEO-II/II.V data set, which dominated $B$ physics for the past decade, amounted to $9\,{\rm fb}^{-1}$.  The $B$ Factories have extremely broad physics programs, but for the central questions of CKM physics they are expected to provide $\sin2\phi_1$ to about $\pm0.05$, $|V_{ub}|$ to about $10\%$, some information on $\phi_3$ from $B\to DK$, $B\to\pi\pi$ or $B\to\rho\pi$, and a great deal of data on rare $B$ decays.

Tevatron Run II is also underway, and CDF and D0 will do exciting $B$ physics.  In particular, they should observe $B_s$ mixing, and measure $\Delta M_{B_s}$ to a percent or so if it is anywhere near its Standard Model value.  They will also measure $\sin2\phi_1$ to about $\pm0.05$, and perhaps provide some information on $\phi_3$ from $B\to DK$ or $B_s\to D_sK$.

Near the end of the decade, the dedicated hadronic $B$ experiments LHCb and BTeV are expected to begin taking data.  Although they will improve the measurements of $\sin2\phi_1$ and $\Delta M_{B_s}$, their greatest value will be to open up rich programs of $B_s$ physics and very rare $B$ decays.  In particular, they should extract $\phi_3$ cleanly from $B_s\to D_sK$, measure CP violating asymmetries in $B_s$ decays, and search for signs of $CP$ violation in $B_s$ mixing.

Finally, one begins to hear discussion of a next generation $e^+e^-$ $B$ Factory, operating at an instantaneous luminosity as high as $10\,{\rm ab}^{-1}{\rm yr}^{-1}$.  Of course, it is still unclear whether such a machine could be built, not to mention a detector which would function in that environment.  But the very idea is exciting, and spurs one to think of  what might be possible with such a facility.  Although one component of a Super $B$ Factory program would be to improve on the analyses performed at BaBar and Belle, I believe that a next generation machine should be justified by its ability to do something which is truly innovative.  Valuable new capabilities might include the collection large samples of CP-tagged $B$ decays~\cite{Falk:2001pd}, or the study of $B_s$ pairs coherently produced at the $\Upsilon(5S)$~\cite{Falk:2000ga}.  Whether to pursue this exciting and challenging direction is the next major question facing our field.

The quest to understand the nature of quark flavor is a long journey which is already more than half a century old.  We have come remarkably far in that time, to the point that we have an elegant theoretical framework whose plausibility has been verified.  We are now in the process of confirming the details of that structure with precision.  Yet we must hope that flaws in our beautiful picture soon will be revealed, leaving us with the new puzzle of what lies beyond.

\section*{Acknowledgements}
It is a pleasure to thank the Organizing Committee for the opportunity to attend this conference, especially at such an exciting (and, sadly, also difficult) time for Japanese high energy physics.  Support was provided by the United States National Science Foundation under Grant PHY--9970781.  A.F.~is a Cottrell Scholar of the Research Corporation.

\end{document}